\documentstyle[12pt]{article}

\def\hybrid{\topmargin -20pt	\oddsidemargin 0pt
	\headheight 0pt	\headsep 0pt
	\textwidth 6.25in	
	\textheight 9.5in	
	\marginparwidth .875in
	\parskip 5pt plus 1pt	\jot = 1.5ex}

\hybrid

\def\baselinestretch{1.2}

\catcode`\@=11

\def\marginnote#1{}
\newcount\hour
\newcount\minute
\newtoks\amorpm
\hour=\time\divide\hour by60
\minute=\time{\multiply\hour by60 \global\advance\minute by-\hour}
\edef\standardtime{{\ifnum\hour<12 \global\amorpm={am}%
	\else\global\amorpm={pm}\advance\hour by-12 \fi
	\ifnum\hour=0 \hour=12 \fi
	\number\hour:\ifnum\minute<10 0\fi\number\minute\the\amorpm}}
\edef\militarytime{\number\hour:\ifnum\minute<10 0\fi\number\minute}

\def\draftlabel#1{{\@bsphack\if@filesw {\let\thepage\relax
   \xdef\@gtempa{\write\@auxout{\string
      \newlabel{#1}{{\@currentlabel}{\thepage}}}}}\@gtempa
   \if@nobreak \ifvmode\nobreak\fi\fi\fi\@esphack}
	\gdef\@eqnlabel{#1}}
\def\@eqnlabel{}
\def\@vacuum{}
\def\draftmarginnote#1{\marginpar{\raggedright\scriptsize\tt#1}}

\def\draft{\oddsidemargin -.5truein
	\def\@oddfoot{\sl preliminary draft \hfil
	\rm\thepage\hfil\sl\today\quad\militarytime}
	\let\@evenfoot\@oddfoot	\overfullrule 3pt
	\let\label=\draftlabel
	\let\marginnote=\draftmarginnote
   \def\@eqnnum{(\theequation)\rlap{\kern\marginparsep\tt\@eqnlabel}%
\global\let\@eqnlabel\@vacuum}  }

\def\preprint{\twocolumn\sloppy\flushbottom\parindent 2em
	\leftmargini 2em\leftmarginv .5em\leftmarginvi .5em
	\oddsidemargin -.5in	\evensidemargin -.5in
	\columnsep .4in	\footheight 0pt
	\textwidth 10.in	\topmargin  -.4in
	\headheight 12pt \topskip .4in
	\textheight 6.9in \footskip 0pt
	\def\@oddhead{\thepage\hfil\addtocounter{page}{1}\thepage}
	\let\@evenhead\@oddhead	\def\@oddfoot{}	\def\@evenfoot{} }

\def\numberbysection{\@addtoreset{equation}{section}
	\def\theequation{\thesection.\arabic{equation}}}

\def\underline#1{\relax\ifmmode\@@underline#1\else
	$\@@underline{\hbox{#1}}$\relax\fi}

\def\titlepage{\@restonecolfalse\if@twocolumn\@restonecoltrue\onecolumn
     \else \newpage \fi \thispagestyle{empty}\c@page\z@
	\def\thefootnote{\fnsymbol{footnote}} }

\def\endtitlepage{\if@restonecol\twocolumn \else \newpage \fi
	\def\thefootnote{\arabic{footnote}}
	\setcounter{footnote}{0}}  

\catcode`@=12
\relax

\def\figcap{\section*{Figure Captions\markboth
	{FIGURECAPTIONS}{FIGURECAPTIONS}}\list
	{Figure \arabic{enumi}:\hfill}{\settowidth\labelwidth{Figure
999:}
	\leftmargin\labelwidth
	\advance\leftmargin\labelsep\usecounter{enumi}}}
 \relax
\def\tablecap{\section*{Table Captions\markboth
	{TABLECAPTIONS}{TABLECAPTIONS}}\list
	{Table \arabic{enumi}:\hfill}{\settowidth\labelwidth{Table
999:}
	\leftmargin\labelwidth
	\advance\leftmargin\labelsep\usecounter{enumi}}}
 \relax
\def\reflist{\section*{References\markboth
	{REFLIST}{REFLIST}}\list
	{[\arabic{enumi}]\hfill}{\settowidth\labelwidth{[999]}
	\leftmargin\labelwidth
	\advance\leftmargin\labelsep\usecounter{enumi}}}
 \relax
\makeatletter
\newcounter{pubctr}
\def\publist{\@ifnextchar[{\@publist}{\@@publist}}
\def\@publist[#1]{\list
	{[\arabic{pubctr}]\hfill}{\settowidth\labelwidth{[999]}
	\leftmargin\labelwidth
	\advance\leftmargin\labelsep
	\@nmbrlisttrue\def\@listctr{pubctr}
	\setcounter{pubctr}{#1}\addtocounter{pubctr}{-1}}}
\def\@@publist{\list
	{[\arabic{pubctr}]\hfill}{\settowidth\labelwidth{[999]}
	\leftmargin\labelwidth
	\advance\leftmargin\labelsep
	\@nmbrlisttrue\def\@listctr{pubctr}}}
 \relax
\makeatother
\newskip\humongous \humongous=0pt plus 1000pt minus 1000pt

\newif\ifdtup

\relax

\def\thefootnote{\fnsymbol{footnote}}
\def\be{\begin{equation}}
\def\ee{\end{equation}}
\def\ba{\begin{eqnarray}}
\def\ea{\end{eqnarray}}

\begin{document}
\renewcommand{\theequation}{\arabic{equation}}
\newcommand{\beq}{\begin{equation}}
\newcommand{\eeq}[1]{\label{#1}\end{equation}}
\newcommand{\ber}{\begin{eqnarray}}
\newcommand{\eer}[1]{\label{#1}\end{eqnarray}}
\begin{titlepage}
\begin{center}

\hfill CERN--TH/95--16\\
\hfill THU--95/01\\
\hfill hep-th/9502065\\

\vskip .3in

{\large \bf T--DUALITY AND WORLD--SHEET SUPERSYMMETRY}

\vskip 0.4in

{\bf Ioannis Bakas}
\footnote{Permanent address: Department of Physics, University of
Patras,
GR--26110 Patras, Greece}
\footnote{e--mail address: BAKAS@SURYA11.CERN.CH}\\
\vskip .1in

{\em Theory Division, CERN\\
     CH-1211 Geneva 23, Switzerland}\\

\vskip .2in

and

\vskip .2in

{\bf Konstadinos Sfetsos}
\footnote{e--mail address: SFETSOS@FYS.RUU.NL}\\
\vskip .1in

{\em Institute for Theoretical Physics, Utrecht University\\
     Princetonplein 5, TA 3508, The Netherlands}\\

\vskip .1in

\end{center}

\vskip .5in

\begin{center} {\bf ABSTRACT } \end{center}
\begin{quotation}\noindent
Four--dimensional string backgrounds with local realizations of
$N=4$ world--sheet supersymmetry
have, in the presence of a rotational Killing symmetry, only one
complex structure which is an $SO(2)$ singlet, while the other two
form an $SO(2)$ doublet. Although $N=2$ world--sheet supersymmetry
is always preserved under Abelian T--duality transformations,
$N=4$ breaks down to $N=2$ in the rotational case. A non--local
realization of $N=4$ supersymmetry emerges, instead, with
world--sheet parafermions. For $SO(3)$--invariant metrics of purely
rotational type, like the Taub--NUT and the Atiyah--Hitchin metrics,
none of the locally realized extended world--sheet supersymmetries
can be preserved under non--Abelian duality.

\end{quotation}
\vskip .2cm
CERN--TH/95--16\\
THU--95/01\\
February 1995\\
\end{titlepage}
\vfill
\eject

\def\baselinestretch{1.2}
\baselineskip 16 pt
\noindent
The duality symmetries of string theory describe a quantum
equivalence between the underlying conformal field theories
of various backgrounds with different geometrical (or even
topological) properties. The target space duality between string
backgrounds with Abelian isometries is the most notable
example, known as T--duality (see, for instance, [1] and references
therein). The toroidal compactification
of the heterotic string theory seems to exhibit another
remarkable symmetry under an $SL(2, Z)$ group of transformations
that act on the coupling constant of the theory. This is the
S--duality and it is more conveniently described as a symmetry
of the axion--dilaton system in the Einstein frame of string
theory [2, 3]. It is natural, once both types of duality symmetries
exist in a string theory, to intertwine them in order to construct
new discrete symmetries [4, 5, 6]. It has recently been
recognized, however, that such a procedure is not always compatible
with the $N=4$ world--sheet supersymmetry of superconformal
string vacua [6]. The condition that characterizes the occurrence of
such an obstruction was linked to the nature of the
Abelian isometries of the corresponding string vacua.

The purpose of the present work is to provide a natural explanation
of this obstruction by considering the explicit behaviour of the
locally realized extended world--sheet supersymmetries under generic
T--duality transformations. As we will see later,
the local realizations of $N=4$ world--sheet supersymmetry cannot be
preserved by
performing T--duality
transformations with respect to rotational Killing vector fields,
due to the peculiar behaviour of the three underlying
independent complex structures. This problem does not arise
for translational Killing symmetries.
Hence, the standard arguments
that relate the $N=4$ world--sheet supersymmetry to the geometrical
properties of the target space
manifold will not be applicable after dualizing
with respect to rotational Killing symmetries.
The rotational Killing symmetries also
spoil the standard supersymmetry transformations in the target space
after T--duality [6, 7], but we will not
elaborate on this issue. We will only remark at the end that the
space--time supersymmetry may be realized differently in terms of
the new target space fields after
a rotational T--duality transformation.
The question we are addressing here
is also interesting for exploring the supersymmetric properties of
the
$O(d,d)$ deformations of a given conformal field theory background,
in general, and in particular along the $J \bar{J}$--line of marginal
deformations [8] (see also [9]).

The behaviour of extended world--sheet supersymmetry under T--duality
transformations is important for understanding the sense in which
target space dualities are string symmetries. It is true that the
$N=2$
world--sheet supersymmetry remains local under generic Abelian
T--duality
transformations. This problem was investigated in the literature
before
by constructing explicitly the relevant complex structure in the dual
formulation (see [10] and references therein).
On target spaces with torsion, there are two complex
structures, a left and a right one, which are relevant for $N=2$
supersymmetry. When the torsion is zero the left and the right
complex structures are identical, but they transform differently
under
Abelian T--duality transformations. This issue and the properties of
their
commutator have already been considered in the most general case and
we will
not deal with them further.
Our question is more elementary and refers to the
local properties of the three independent complex structures
(either the left or
the right) in superconformal theories with $N=4$ supersymmetry.
Such complications
are certainly not present in theories with only $N=2$ supersymmetry.

The target space metric in $N=4$ superconformal theories with torsion
is constrained to satisfy the conditions
\be
g = \Omega g^{\prime} ~ , ~~~~ {\Box}^{\prime} \Omega = 0 ~ ,
\ee
where $g^{\prime}$ is a hyper--Kahler metric [11] (see also [12]).
This theorem is true provided
that the three underlying complex structures are locally realized, in
which
case they are covariantly constant (including torsion). Recall
that there are examples of $N=4$, $\hat{c} = 4$ models where the
above
geometric conditions are not satisfied; in all such models, the $N=4$
supersymmetry is non--locally realized on the world--sheet [13]. We
will
demonstrate that T--duality with respect to rotational Killing
symmetries
always leads to such a peculiar world--sheet behaviour. For example,
the
string background which is dual to flat space with respect to any
one of its rotational
Killing symmetries can be easily shown to contradict the conditions
(1).
It is also conceivable that all models with non--local realizations
of $N=4$ supersymmetry
may admit a local description, in an appropriately chosen T--dual
formulation,
with respect to some rotational Killing symmetry.

In the following, we simplify our presentation by considering first
the supersymmetric behaviour of pure gravitational string backgrounds
under
generic T--duality transformations and review some of the basic
concepts.
The geometric description of the Abelian duality in terms of
canonical
transformations in the target space [14, 15] will be particularly
useful in
finding the explicit form of the dual complex structures. We will
also
discuss these issues for some more general string backgrounds,
including
the worm--hole solution [12, 16]. Finally we will conclude with some
generic
features of non--Abelian duality and supersymmetry.

Four--dimensional pure gravitational backgrounds with $N=4$
world--sheet supersymmetry are known to be hyper--Kahler
manifolds (i.e. $\Omega =1$ in eq. (1)).
Their Riemann tensor satisfies the self--duality
conditions
\be
R_{\mu \nu \rho \sigma} = \pm {1 \over 2} \sqrt{\det G} ~
{{\epsilon}_{\rho \sigma}}^{\kappa \lambda}
R_{\mu \nu \kappa \lambda}
\ee
with the $\pm$ sign corresponding to self--dual or anti--self--dual
metrics respectively, depending on the conventions. If these
backgrounds also admit (at least) one Killing symmetry, then, in the
special coordinate system where the Killing symmetry becomes
manifest,
their metric will take a particularly simple form. There are two
kinds
of Killing symmetries that are in fact distinct from each other.
The first kind, which is usually called translational, corresponds to
Killing vector fields $K_{\nu}$ with self--dual covariant
derivatives,
\be
{\nabla}_{\mu} K_{\nu} = \pm {1 \over 2} \sqrt{\det G} ~
{{\epsilon}_{\mu \nu}}^{\rho \sigma} {\nabla}_{\rho} K_{\sigma} ~,
\ee
according to the two cases (2) respectively [17]. The translational
symmetries
are also known as triholomorphic, because of their special
character in Kahlerian coordinates. The second kind,
which is usually called rotational, encompasses all other Killing
vector fields. It is true that rotational symmetries are more rare
than translational ones, although simple examples of self--dual
metrics
admit both; these include the flat space, the Taub--NUT and the
Eguchi--Hanson gravitational instanton.

In the case of 4--dim hyper--Kahler manifolds with a translational
Killing
symmetry, the metric assumes the form
\be
ds^{2} = V(dT + {\Omega}_{1} dX + {\Omega}_{2} dY + {\Omega}_{3}
dZ)^{2}
+ V^{-1} (dX^{2} + dY^{2} + dZ^{2}) ~,
\ee
where $T$ is a coordinate adapted for the translational Killing
vector
field $\partial / \partial T$. Moreover, ${\Omega}_{i}$ are
constrained
to satisfy the special conditions
\be
{\partial}_{i} V^{-1} = \pm {1 \over 2} {\epsilon}_{ijk}
({\partial}_{j}{\Omega}_{k} - {\partial}_{k} {\Omega}_{j}) ~ ,
\ee
depending on the self--dual or the anti--self--dual character of the
metric
respectively [18]. It also follows that $V^{-1}$ satisfies the 3--dim
flat space
Laplace equation. Localized solutions of this equation correspond to
the
familiar series of multi--centre Eguchi--Hanson gravitational
instantons
or to the multi--Taub--NUT family, depending on the asymptotic
conditions
on $V^{-1}$ (see, for instance, [19] and references therein).

The three independent complex structures associated to self--dual
metrics
with translational symmetry have been explicitly constructed in the
literature [20]. In the special coordinate system (4) they assume the
particularly simple form
\ba
F_{1} & = & (dT + {\Omega}_{2} dY) \wedge dX - V^{-1} dY \wedge dZ ~,
\\
F_{2} & = & (dT + {\Omega}_{1} dX) \wedge dY + V^{-1} dX \wedge dZ ~,
\\
F_{3} & = & (dT + {\Omega}_{1} dX + {\Omega}_{2} dY) \wedge dZ -
V^{-1} dX \wedge dY ~,
\ea
where ${\Omega}_{3}$ has been set equal to zero by a gauge
transformation.
It is straightforward to verify that they are covariantly constant
on--shell and satisfy the $SU(2)$ Clifford algebra. The property that
is important for our purposes is that all three complex structures
remain
invariant under $T$--shifts. This will be crucial for understanding
the
way the local realizations of $N=4$ world--sheet supersymmetry
behave, in a string setting, under T--duality transformations. It is
also useful to note at this point that $S_{\pm} = b \pm V$, where $b$
is
the nut potential of the metric (4), is constant for self--dual or
anti--self--dual metrics respectively [6].

In the case of 4--dim hyper--Kahler manifolds with a rotational
Killing symmetry, there exists a coordinate system ($\tau$, $x$, $y$,
$z$)
in which the corresponding line element takes the form
\be
ds^2 = v(d \tau + {\omega}_{1} dx + {\omega}_{2} dy)^{2} +
v^{-1} \left(e^{\Psi} {dx}^{2} + e^{\Psi} {dy}^{2} + {dz}^{2}\right).
\ee
In these adapted coordinates the rotational Killing vector field is
$\partial / \partial \tau$ and all the components of the metric are
expressed in terms of a single scalar field $\Psi (x, y, z)$ [17], so
that
\be
v^{-1} = {\partial}_{z} \Psi ~ , ~~~~ {\omega}_{1} = \mp
{\partial}_{y} \Psi ~,
{}~~~ {\omega}_{2} = \pm {\partial}_{x} \Psi ~ ,
\ee
where $\Psi (x, y, z)$ satisfies the continual Toda equation:
\be
({{\partial}_{x}}^{2} + {{\partial}_{y}}^{2}) \Psi +
{{\partial}_{z}}^{2} e^{\Psi} = 0 ~.
\ee
These coordinates actually provide the geodesic form of the
corresponding
3--dim line element for which $S_{\pm} = b \pm v = - z$, and hence is
not constant [6].
We use capital (small) letters to distinguish the special coordinates
in
the presence of  translational (rotational) Killing symmetries.

Metrics with rotational Killing symmetry differ from those with
translational symmetry
in that not all three independent complex structures can be chosen to
be
$\tau$--shift invariant. In fact, only one complex structure can be
chosen to
be an $SO(2)$ singlet, while the other two necessarily form an
$SO(2)$ doublet.
We have explicitly
\be
F_{3} = (d \tau + {\omega}_{1} dx + {\omega}_{2} dy) \wedge dz +
v^{-1} e^{\Psi} dx \wedge dy
\ee
for the singlet and
\be
\left( \begin{array}{c}
F_{1}\\
     \\
F_{2} \end{array} \right) = e^{{1 \over 2} \Psi}
\left( \begin{array}{cr}
\cos {\tau \over 2} & \sin {\tau \over 2} \\
                    &                     \\
\sin {\tau \over 2} & - \cos {\tau \over 2} \end{array} \right)
\left( \begin{array}{c}
f_{1}\\
     \\
f_{2} \end{array} \right)
\ee
for the doublet, where
\be
f_{1} = (d \tau + {\omega}_{2} dy) \wedge dx - v^{-1} dz \wedge dy ~,
\ee
\be
f_{2} = (d \tau + {\omega}_{1} dx) \wedge dy + v^{-1} dz \wedge dx ~.
\ee
It can be easily verified, once the right guess has been made, that
$F_{1}$, $F_{2}$ and $F_{3}$ are covariantly constant on--shell (11)
and
satisfy the $SU(2)$ Clifford algebra
\footnote{It is known that the continual Toda equation (11) exhibits
a $W_{\infty}$ symmetry on--shell [21]. This symmetry preserves the
sphere of complex structures defined by $F_{1}$, $F_{2}$ and
$F_{3}$.}.
The general explicit construction of all
three complex structures in the rotational frame (9) has not appeared
in the literature before, to the best of our knowledge. The
expressions
(12)--(15) are strictly speaking correct only for self--dual metrics.
Their anti--self--dual counterparts, together with the metric (9),
can
simply be obtained using the substitution $\tau \rightarrow - \tau$.

It is possible to guess the general form of the three complex
structures
in the rotational frame by considering the special example of the
Eguchi--Hanson gravitational instanton. This background is
$SO(3)$--symmetric, and with respect to one of its three
translational
Killing symmetries it can be written in the form (4), where
\be
V^{-1} = {1 \over R_{+}} + {1 \over R_{-}} ~ , ~~~
{\Omega}_{1} = - {Y \over X^2 + Y^2} \left({Z_{+} \over R_{+}} +
{Z_{-} \over R_{-}}\right) , ~~~
{\Omega}_{2} = {X \over X^2 + Y^2} \left({Z_{+} \over R_{+}} +
{Z_{-} \over R_{-}}\right) ,
\ee
with ${\Omega}_{3} = 0$ and
\be
Z_{\pm} = Z \pm a ~ , ~~~~ R_{\pm}^{2} = X^2 + Y^2 + Z_{\pm}^{2} ~.
\ee
The moduli parameter is $a$. The Eguchi--Hanson instanton also has an
additional $U(1)$ Killing symmetry which is rotational. The
coordinate
transformation that makes the latter manifest is
\be
\left( \begin{array}{c}
X\\
     \\
Y \end{array} \right) = e^{{1 \over 2} \Psi}
\left( \begin{array}{cr}
\cos {\tau \over 2} & \sin {\tau \over 2} \\
                    &                     \\
\sin {\tau \over 2} & - \cos {\tau \over 2} \end{array} \right)
\left( \begin{array}{c}
x\\
     \\
y \end{array} \right), ~~~
Z= z ~ {1 - {1 \over 8} (x^2 + y^2) \over 1 + {1 \over 8} (x^2 +
y^2)} ~ ,
{}~~~ T = 2 ~ {\tan}^{-1} {y \over x} ~ ,
\ee
where
\be
\Psi (x, y, z) = \log {z^2 - a^2 \over 2 \left(1 + {1 \over 8} (x^2 +
y^2)
\right)^{2}}
\ee
is the corresponding Toda potential. It follows that the three
independent
complex structures (6)--(8) are mapped to (12)--(15) by the
coordinate
transformation described above. Similar remarks apply to the flat
space metric,
which has $\Psi (x, y, z) = \log z$ in the rotational frame (9).

It is useful to recall, before we proceed further, that the Taub--NUT
and the
Atiyah--Hitchin metric on the moduli space of the $SU(2)$ 2--monopole
solutions
(in the BPS limit) [22] provide another two non--trivial examples of
4--dim
hyper--Kahler manifolds with rotational Killing symmetries. Both of
them
admit an $SO(3)$ isometry of purely rotational type, but the
Taub--NUT
metric also admits an additional $U(1)$ symmetry which is
translational.
(This should be compared with the opposite character of the
Eguchi--Hanson
Killing symmetries.) Moreover, this completes the classification of
the complete
$SO(3)$--invariant self--dual metrics [20]. The Atiyah--Hitchin
metric can be
regarded as the simplest example of hyper--Kahler manifolds with only
rotational Killing symmetries. Since the existence of only two
rotational
symmetries (and no other of either type) is not allowed from general
considerations [17], it follows that the next more complicated
solutions
of this kind
(if they exist at all) will exhibit only one rotational Killing
symmetry
and no other symmetries of either type.
Work in this direction is in progress, while determining
the Toda potential that corresponds to the Atiyah--Hitchin metric,
its
possible generalization
to a new series of rotational instantons, and its
use as a supersymmetric gravitational string background [23].

We have already seen that, in metrics with (at least) one Killing
symmetry,
either all complex structures behave as singlets or one of
them is a singlet and the remaining two form an $SO(2)$ doublet.
This applies to the conformal class of 4--dim metrics (1) and we
will next investigate the effect of the T--duality on them.
We first recall the essential ingredients for describing the Abelian
T--duality transformation as a canonical transformation in the
target space [15] (see also [14] for some earlier ideas).
The classical propagation of strings in a general target space with
metric
$G_{\mu \nu}(X)$ and antisymmetric tensor field $B_{\mu \nu}(X)$ is
described by the 2--dim $\sigma$--model density
$Q_{\mu \nu} \partial X^{\mu} \bar{\partial} X^{\nu}$, where
\be
Q_{\mu \nu} = G_{\mu \nu} + B_{\mu \nu} ~.
\ee
Consider backgrounds with a Killing symmetry generated by the vector
field $\partial / \partial X^{0}$ ($X^{0}$ will be $T$ or $\tau$ in
the
translational or in the rotational frame, respectively). Let $P_{0}$
be the
canonical momentum conjugate to the string variable $X^{0}$; we
perform the canonical transformation ($X^{0}$, $P_{0}$) $\rightarrow$
(${\tilde{X}}^{0}$, ${\tilde{P}}_{0}$), as is prescribed by the
interchange $P_{0} \leftrightarrow {\partial}_{\sigma} X^{0}$, where
$\sigma$ denotes the spatial coordinate on the string world--sheet.
This is equivalent to the transformation
\be
\partial X^{0} \rightarrow {1 \over G_{00}} (\partial X^{0} -
Q_{i0} \partial X^{i}) ~ , ~~~~
\bar{\partial} X^{0} \rightarrow - {1 \over G_{00}} (\bar{\partial}
X^{0}
+ Q_{0i} \bar{\partial} X^{i}) ~ ,
\ee
using the expression for the canonical momentum $P_{0}$. It is then
straightforward to read--off the form of the dual string background
by
substituting (21) into the Hamiltonian form of the 2--dim action. The
result,
\be
{\tilde{G}}_{00} = {1 \over G_{00}} ~, ~~~ {\tilde{Q}}_{0i} =
{Q_{0i} \over G_{00}} ~, ~~~ {\tilde{Q}}_{i0} = - {Q_{i0} \over
G_{00}}
{}~, ~~~ {\tilde{Q}}_{ij} = Q_{ij} - {Q_{i0} Q_{0j} \over G_{00}} ~,
\ee
indeed describes the Abelian T--duality transformation in all
generality.
The conformal invariance also requires that the corresponding dilaton
field $2 \Phi$ is shifted by $- \log G_{00}$ (see, for instance,
[1]).

The transformation (21) amounts to a non--local redefinition of the
target space variable associated with the Killing symmetry,
\be
X^{0} \rightarrow \int {1 \over G_{00}} \left( (\partial X^{0} -
Q_{i0} \partial X^{i}) d z - (\bar{\partial} X^{0} + Q_{0i}
\bar{\partial} X^{i}) d \bar{z} \right) ~.
\ee
Despite the non--localities, the dual target space fields (22) are
locally related to the original ones. However, other geometrical
quantities in the target space, such as the
Kahler 2--forms describing the complex
structures, are not bound to be always local in the dual picture.
This is
precisely our concern for addressing the question of local versus
non--local
realizations of $N=4$ world--sheet supersymmetry, in general, upon
duality.
We are considering the local behaviour of three independent
almost complex structures, which are actually Hermitian,
without worrying about their integrability, since the Nijenhuis
conditions
are not necessary for the existence of extended world--sheet
supersymmetry
[24].

The complex structures come in pairs in the presence of torsion and
define the 2--forms
\be
F_{I}^{\pm} = {(F_{I}^{\pm})}_{\mu \nu} d X^{\mu} \wedge d X^{\nu} =
2 {(F_{I}^{\pm})}_{0i} d X^{0} \wedge d X^{i} + {(F_{I}^{\pm})}_{ij}
d X^{i} \wedge d X^{j} ~
\ee
that satisfy all the necessary conditions (they are covariantly
constant,
including the torsion,
and each $+$ or $-$ set forms separately an $SU(2)$ Clifford
algebra).
The $F_{I}^{\pm}$ are associated to right or left--handed
fermions and in order to find the correct transformation
properties under T--duality, we simply have to use the replacement
$d X^{\mu} \rightarrow \partial X^{\mu}$ for $F_{I}^{+}$ and
$d X^{\mu} \rightarrow  \bar{\partial} X^{\mu}$ for $F_{I}^{-}$.
Of course, this is only meant to be a prescription for
extracting the relevant part of the complex structures under the
duality transformation (21). With this explanation in mind, we
first consider the simplest case having all ${(F_{I}^{\pm})}_{\mu
\nu}$
independent of $X^{0}$. The result we obtain for the dual complex
structures in component form is
\be
{({\tilde{F}}_{I}^{+})}_{0i} = {1 \over G_{00}} {(F_{I}^{+})}_{0i} ~,
{}~~~
{({\tilde{F}}_{I}^{+})}_{ij} = {(F_{I}^{+})}_{ij} + {1 \over G_{00}}
\left( {(F_{I}^{+})}_{0i} Q_{j0} - {(F_{I}^{+})}_{0j} Q_{i0} \right)
\ee
and
\be
{({\tilde{F}}_{I}^{-})}_{0i} = - {1 \over G_{00}} {(F_{I}^{-})}_{0i}
{}~, ~~~
{({\tilde{F}}_{I}^{-})}_{ij} = {(F_{I}^{-})}_{ij} + {1 \over G_{00}}
\left( {(F_{I}^{-})}_{0i} Q_{0j} - {(F_{I}^{-})}_{0j} Q_{0i} \right)
{}~.
\ee
If we were computing ${{({\tilde{F}}_{I}^{\pm})}^{\mu}}_{\nu} =
{\tilde{G}}^{\mu \lambda} {({\tilde{F}}_{I}^{\pm})}_{\lambda \nu}$ in
this case, the result would coincide with the expressions for the
dual
complex stuctures derived before [10].

It follows from the previous analysis that all the components
${(F_{I}^{\pm})}_{\mu \nu}$ will be independent of $X^{0}$ if the
Killing vector field $\partial / \partial X^{0} \equiv
\partial / \partial T$ is translational. We consider the effect of
T--duality on the self--dual gravitational backgrounds (4), (5) as
an application in this case. The dual background is conformally flat,
\be
d {\tilde{s}}^{2} = V^{-1} (d T^{2} + d X^{2} + d Y^{2} + d Z^{2}) ~,
\ee
with a non--trivial antisymmetric tensor field
\be
\tilde{B} = 2 ~ d T \wedge ({\Omega}_{1} d X + {\Omega}_{2} d Y)
\ee
and dilaton field $2 \tilde{\Phi} = \log V^{-1}$.
The corresponding dual complex
structures can be obtained from eqs. (6)--(8) and they assume the
form
\be
{\tilde{F}}_{I}^{\pm} = V^{-1} \left( \pm d T \wedge d X^{I} -
{1 \over 2} {\epsilon}_{IJK} d X^{J} \wedge d X^{K} \right) ~,
\ee
where $\{ X^{I} \}$ = $\{ X, Y, Z \}$. The dual background is
consistent
with the condition (1), because the conformal factor $\Omega =
V^{-1}$
satisfies the Laplace equation in flat space that was imposed by the
self--duality of the original metric. The resulting backgrounds
are the axionic instantons introduced in the toroidal
compactification
of the heterotic string theory [12, 25]. In this case, the solutions
exhibit $N=4$ world--sheet supersymmetry, which is locally realized
and
hence compatible with the geometric characterization (1) of the
target
space metric before and after duality.

The situation is radically different when the T--duality is performed
with
respect to a rotational Killing symmetry. A generic string background
with
locally realized $N=4$ world--sheet supersymmetry has a hyper--Kahler
metric $g^{\prime}$ associated with it, according to eq. (1). Using
the
rotational frame (9), in which $X^{0} = \tau$, the T--duality
transformation yields the background
\be
d {\tilde{s}}^{2} = v^{-1} (e^{\Psi} d x^{2} + e^{\Psi} d y^{2} +
d z^{2} + d {\tau}^{2}) ~,
\ee
with antisymmetric tensor field
\be
\tilde{B} = 2 ~ d \tau \wedge ({\omega}_{1} dx + {\omega}_{2} dy )
\ee
and dilaton field $2 \tilde{\Phi} = \log v^{-1}$. The complex
structure (12) will
remain local in the dual picture, assuming the form
\be
{\tilde{F}}_{3}^{\pm} = v^{-1} (\pm d \tau \wedge dz + e^{\Psi}
dx \wedge dy ) ~.
\ee
The 2--forms (14), (15) become similarly
\be
{\tilde{f}}_{1}^{\pm} = v^{-1} (\pm d \tau \wedge dx - dz \wedge dy )
{}~ ,
{}~~~
{\tilde{f}}_{2}^{\pm} = v^{-1} (\pm d \tau \wedge dy + dz \wedge dx )
{}~ .
\ee
The duality transformation amounts to the canonical transformation
\be
\tau \rightarrow \int (v^{-1} \partial \tau - {\omega}_{1} \partial x
- {\omega}_{2} \partial y) dz - (v^{-1} \bar{\partial} \tau +
{\omega}_{1} \bar{\partial} x + {\omega}_{2} \bar{\partial} y)
d \bar{z}
\ee
and so the components of the forms $F_{1}$ and $F_{2}$, which depend
explicitly on $\tau$ via trigonometric functions, will become
non--local after the rotational T--duality. Moreover, the resulting
${\tilde{F}}_{1}^{\pm}$ and ${\tilde{F}}_{2}^{\pm}$ are not
covariantly
constant on--shell, including the torsion coming from eq. (31).

We found that the local realization of $N=4$ world--sheet
supersymmetry
breaks down to $N=2$, with ${\tilde{F}}_{3}^{\pm}$ providing the
relevant pair of complex structures in the presence of torsion. This
is
also consistent with the fact that the dual background (30) is
not conformally equivalent to a hyper--Kahler metric, as would have
been required otherwise by $N=4$ world--sheet supersymmetry. It
should
be noted, though, that general arguments from superconformal field
theory indicate that the $N=4$ world--sheet supersymmetry will remain
present, but part of it will become hidden into a non--local
realization. We do not have an exact conformal field theory
description
of the string gravitational background (30), (31) in order to
illustrate this point in all generality. For this reason we will
examine the question in the special case of the 4--dim worm--hole
solution and its rotational dual background, where an exact
description is available in terms of the $SU(2)$ WZW model and its
derivatives. We will see later that the parafermion currents of
the $SU(2)/U(1)$ coset model describe the non--local structure of the
dual 2--forms ${\tilde{F}}_{1}^{\pm}$, ${\tilde{F}}_{2}^{\pm}$, thus
providing the explicit construction of a non--locally realized
$N=4$ superconformal algebra with $\hat{c} = 4$ [13].

An analogous situation
arises when a string background exhibits a non--Abelian symmetry
group.
Various Killing symmetries that do not commute with a rotational
isometry will become non--locally realized after
performing the T--duality. They
all remain symmetries of the dual model, but some of
them are hidden in the non--local
realization of the corresponding symmetry algebra. This issue is
illustrated
with a simple 2--dim example, which we give separately as an
appendix.

The worm--hole solution of 4--dim string theory provides an exact
conformal field theory background with $N=4$
world--sheet supersymmetry [12, 13, 16]. The $N=4$ superconformal
algebra is locally realized in terms of four bosonic currents, three
non--Abelian $SU(2)_{k}$ currents and one Abelian current with
background
charge $Q = \sqrt{2/(k+2)}$, so that the central charge is $\hat{c} =
4$.
There are also four free--fermion superpartners and the solution is
described by the $SU(2)_{k} \bigotimes U(1)_{Q}$ supersymmetric WZW
model
\footnote{The realization of the $N=4$ superconformal algebra in
terms of
$SU(2)$ currents was first considered in ref. [26].}.
The background fields of this model are given in holomorphic target
space coordinates
\ba
d s^{2} & = & V^{-1} (du d \bar{u} + dw d \bar{w}) ~, ~~~~~
V = u \bar{u} + w \bar{w} ~,\\
B & = & {1 \over 2} V^{-1} (u \bar{u} - w \bar{w})
\left( {1 \over u \bar{w}} d u \wedge d \bar{w} + {1 \over \bar{u} w}
d \bar{u} \wedge dw \right) ~,
\ea
with a non--trivial dilaton field $2 \Phi = \log V^{-1}$. This
background
is conformally flat and it satisfies the condition (1) as required.
We note
for completeness that it is not T--dual to a pure gravitational
self--dual
background with respect to a translational Killing symmetry.
We have also determined the three independent pairs of the underlying
complex
structures of the model,
\ba
F_{1}^{+} & = & {1 \over 2} V^{-1} (du \wedge dw + d \bar{u} \wedge d
\bar{w}) ~,
{}~~~
F_{1}^{-} = {i \over 2} V^{-1} (du \wedge d \bar{w} - d \bar{u}
\wedge dw) ~,\\
F_{2}^{+} & = & {i \over 2} V^{-1} (du \wedge dw - d \bar{u} \wedge d
\bar{w}) ~,
{}~~~
F_{2}^{-} = {1 \over 2} V^{-1} (du \wedge d \bar{w} + d \bar{u}
\wedge dw) ~,\\
F_{3}^{+} & = & {i \over 2} V^{-1} (du \wedge d \bar{u} + dw \wedge d
\bar{w}) ~,
{}~~~
F_{3}^{-} = {i \over 2} V^{-1} (-du \wedge d \bar{u} + dw \wedge d
\bar{w}) ~,
\ea
which satisfy all the necessary conditions for having $N=4$
supersymmetry.

It is known that the worm--hole solution is dual to the exact
superconformal
field theory based on the WZW model
$SU(2)_{k}/U(1) \bigotimes U(1) \bigotimes U(1)_{Q}$ with $\hat{c} =
4$
[10, 13] (see also [27]).
This can be demonstrated by introducing the polar coordinates $\rho$,
$\tau$, $\psi$ and $\varphi$ in the target space, so that
\be
u = e^{\rho + i \tau} \cos \varphi ~, ~~~~~
w = e^{\rho + i \psi} \sin \varphi ~,
\ee
and performing the T--duality transformation in the rotational
$\tau$ direction. It is actually more convenient
for calculational purposes to consider the
target space variables
\be
\alpha = {1 \over 2} \psi - \tau ~, ~~~~~
\beta = {1 \over 2} \psi + \tau ~.
\ee
Then, the duality transformation reads
\be
\tau \rightarrow \tilde{\tau} \equiv
\int (\partial \beta - {\tan}^{2} \varphi ~
\partial \alpha) dz - (\bar{\partial} \beta - {\tan}^{2} \varphi ~
\bar{\partial} \alpha) d \bar{z}
\ee
and the resulting new background has no
antisymmetric tensor field. The dual metric assumes the form
\be
d {\tilde{s}}^{2} = d {\rho}^{2} + d {\beta}^{2} + d {\varphi}^{2}
+ {\tan}^{2} \varphi ~ d {\alpha}^{2} ~,
\ee
while the corresponding dilaton field is $- 2 \tilde{\Phi} = 2 \rho +
\log ({\cos}^{2} \varphi)$. This string background, which corresponds
to the $SU(2)/U(1)_{k} \bigotimes U(1) \bigotimes U(1)_{Q}$ coset
model,
has the special feature that the metric (43) is not hyper--Kahler.
Both
conformal field theories admit an $N=4$ superconformal symmetry, but
it
is not surprising that it is non--locally realized in the latter.

We will demonstrate that this non--local realization of the
$N=4$ world--sheet supersymmetry provides a simple example of our
general framework. The complex structure, which is dual to (39),
turns
out to be local:
\be
{\tilde{F}}_{3} = d \rho \wedge d \beta + \tan \varphi ~
d \varphi \wedge d \alpha ~,
\ee
and there is no distinction between the $+$ and $-$ components, since
$\tilde{B} = 0$. The complex structures (37) and (38), on the other
hand, become non--locally realized in the dual model. The resulting
expressions, up to an irrelevant factor of $1/2$, are
\ba
{\tilde{F}}_{1}^{+} & = & (d \rho + i ~ d \beta) \wedge {\Psi}_{+}
+ (d \rho - i ~ d \beta) \wedge {\Psi}_{-} ~ ,\\
{\tilde{F}}_{1}^{-} & = & i (d \rho - i ~ d \beta) \wedge
{\bar{\Psi}}_{+}
- i (d \rho + i ~ d \beta) \wedge {\bar{\Psi}}_{-}
\ea
and
\ba
{\tilde{F}}_{2}^{+} & = & i (d \rho + i ~ d \beta) \wedge {\Psi}_{+}
- i (d \rho - i ~ d \beta) \wedge {\Psi}_{-} ~ ,\\
{\tilde{F}}_{2}^{-} & = & (d \rho - i ~ d \beta) \wedge
{\bar{\Psi}}_{+}
+ (d \rho + i ~ d \beta) \wedge {\bar{\Psi}}_{-} ~,
\ea
where
\be
{\Psi}_{\pm} = (d \varphi \pm i \tan \varphi ~ d \alpha)
e^{\pm i (\tilde{\tau} + \alpha + \beta)} ~, ~~~~
{\bar{\Psi}}_{\pm} = (d \varphi \mp i \tan \varphi ~ d \alpha)
e^{\pm i (\tilde{\tau} - \alpha - \beta)}
\ee
are non--local 1--forms with $\tilde{\tau}$ given by eq. (42).
The above 2--forms are not covariantly constant on--shell and they
are
responsible for having a non--local realization of the $N=4$
world--sheet
supersymmetry in the dual to the worm--hole string model [13].

The non--local 1--forms (49) can be
naturally decomposed into (1, 0) and (0, 1)
forms on the string world--sheet,
\be
{\Psi}_{\pm} = {\Psi}_{\pm}^{(1, 0)} dz + {\Psi}_{\pm}^{(0, 1)} d
\bar{z}
{}~, ~~~~
{\bar{\Psi}}_{\pm} = {\bar{\Psi}}_{\pm}^{(1, 0)} dz +
{\bar{\Psi}}_{\pm}^{(0, 1)} d \bar{z} ~.
\ee
It can be easily verified that
\be
\bar{\partial} {\Psi}_{\pm}^{(1, 0)} = 0 ~, ~~~~~
\partial {\bar{\Psi}}_{\pm}^{(0, 1)} = 0
\ee
are chirally conserved, using the classical equations of motion of
the
dual model, and in fact ${\Psi}_{\pm}^{(1, 0)}$ and
${\bar{\Psi}}_{\pm}^{(0, 1)}$ coincide with the classical
parafermions
that exist in this case; the field $\beta$ actually provides the
dressing
of the $SU(2)/U(1)$ parafermions to the full 4--dim coset model.
Hence, the
usual $N=4$ world--sheet supersymmetry of
the 4--dim worm--hole background breaks down to $N=2$ and a
non--local
realization of $N=4$ emerges instead in the dual model with
world--sheet
parafermions. The T--duality transformation with respect to a
rotational
Killing symmetry is clearly the reason for this behaviour in string
theory.

We conclude with a few remarks concerning the effect of the
non--Abelian
duality transformations on gravitational backgrounds with extended
world--sheet supersymmetry.
For $SO(3)$--invariant metrics, the complex structures either can be
$SO(3)$ singlets, thus remaining invariant under the non--Abelian
group
action,
\be
{\pounds}_{J} {(F_{I})}_{\mu \nu} = 0 ~,
\ee
or form an $SO(3)$ triplet when
\be
{\pounds}_{J} {(F_{I})}_{\mu \nu} = {\epsilon}_{JIK} {(F_{K})}_{\mu
\nu}.
\ee
The Lie derivative is taken with respect to the $SO(3)$ generators of
the
isometry. The Eguchi--Hanson metric corresponds
to the first case, while the Taub--NUT and the Atiyah--Hitchin
metrics to
the second [20]. Although there is no
systematic formulation of the non--Abelian
duality in terms of canonical transformations in the target space, we
can easily find the dual complex structures for backgrounds where the
original complex structures are $SO(3)$ singlets. In this case, the
complex structures are written in terms of the left--invariant
Maurer--Cartan forms of $SO(3)$. In the dual model they are obtained
by replacing ordinary derivatives by covariant ones, using gauge
fields,
and then substituting the on--shell solution for the gauge fields and
fixing a unitary gauge. Such an algorithm will certainly not be
applicable
if the complex structures form an $SO(3)$ triplet. According to this,
the dual version of the Eguchi--Hanson instanton with respect to
$SO(3)$
will have an $N=4$ world--sheet supersymmetry locally realized. In
fact,
the non--Abelian duality has already been performed in this case,
producing a conformally flat metric that satisfies the condition (1)
[28].
On the other hand, performing the non--Abelian $SO(3)$--duality
to the Taub--NUT
and the Atiyah--Hitchin metrics will result in a total loss of all
the
locally realized extended world--sheet supersymmetries. We expect to
have a non--local realization of supersymmetry in such cases with
non--Abelian parafermions. Of course, the $N=1$ supersymmetry will
not
be affected by any kind of T--duality transformations. Clearly, the
relation of non--abelian duality with supersymmetry deserves a more
thorough study.

In summary, we found that the T--duality transformations with respect
to rotational Killing symmetries always break the local realizations
of
the $N=4$ world--sheet supersymmetry to $N=2$ (or even to $N=1$ in
the
non--Abelian case). The effect of the intertwined T--S--T duality
can be even more severe. For example, the T--S--T
transformation of a purely gravitational background with $N=4$
world--sheet
supersymmetry will always yield a purely gravitational metric, which
is
Ricci--flat, but not Kahler, if T is rotational [6]. In this case,
all the
complex structures of the target spece manifold will be destroyed and
therefore, no space--time supersymmetry can exist in the usual sense.
These issues raise the question whether some new space--time
supersymmetry generators can be defined by intertwining the standard
ones with the rotational T--duality transformations,
\be
{\tilde{Q}}_{i} = [T, ~ Q_{i}] ~.
\ee
If this is possible, the ${\tilde{Q}}_{i}$ will have a very different
realization in terms of the new background fields. The main example
for
this investigation is again provided by the coset model
$SU(2)/U(1)_{k} \bigotimes U(1) \bigotimes U(1)_{Q}$. We hope to
return to
this problem elsewhere.

\vskip1cm
\centerline{\bf Acknowledgements}
\vskip.2cm
\noindent
We thank Luis Alvarez--Gaume, Bernard de Wit, Michael Faux,
Fawad Hassan, Elias Kiritsis and Costas
Kounnas for many useful discussions during the course of this work.

\newpage
\centerline{\bf APPENDIX}
\vskip.2cm
\setcounter{section}{0}
\setcounter{equation}{0}
\renewcommand{\theequation}{A.\arabic{equation}}
\noindent
In this appendix we illustrate with a simple example the non--local
realization of Killing symmetries after performing a rotational
T--duality transformation. Similar ideas were presented by Kiritsis
in
the context of parafermionic symmetries [8] (see also [15]).

Consider the 2--dim action of two free bosons written in polar
coordinates
$\rho$, $\varphi$,
\be
S = \int \partial \rho \bar{\partial} \rho + {\rho}^{2}
\partial \varphi \bar{\partial} \varphi ~.
\ee
This action exhibits isometries associated with the following
variations,
\be
\delta \varphi = {\epsilon}_{0} + {\epsilon}_{+} {1 \over \rho}
e^{i \varphi} + {\epsilon}_{-} {1 \over \rho} e^{-i \varphi} ~ , ~~~~
\delta \rho = -i {\epsilon}_{+} e^{i \varphi} + i {\epsilon}_{-}
e^{-i \varphi} ~,
\ee
where ${\epsilon}_{0}$ and ${\epsilon}_{\pm}$ are constant. It can be
verified that the
corresponding vector fields
\be
J_{0} = i {\partial}_{\varphi} ~, ~~~~ J_{\pm} = e^{\mp i \varphi}
\left({1 \over \rho} {\partial}_{\varphi} \pm i {\partial}_{\rho}
\right)
\ee
generate the Euclidean group in two dimensions, i.e.
$[J_{0} , ~ J_{\pm}] = \pm J_{\pm}$ and $[J_{+} , J_{-}] = 0$.

The dual action with respect to the rotational Killing vector field
$J_{0}$ is
\be
\tilde{S} = \int \partial \rho \bar{\partial} \rho + {1 \over
{\rho}^{2}}
\partial \varphi \bar{\partial} \varphi ~,
\ee
while $\varphi$ itself transforms non--locally under T--duality,
\be
\varphi \rightarrow \int {1 \over {\rho}^{2}} (d z \partial
\varphi - d \bar{z} \bar{\partial} \varphi) ~.
\ee
The conformal invariance of the model determines the corresponding
dilaton field, which is irrelevant for the present purposes.
According to this,
$J_{\pm}$ become non--locally realized in the dual formulation
of our toy model. Only $J_{0}$ remains local in this case, generating
with $J_{\pm}$ the same symmetry algebra as before.

The 2--dim theory (A.1) describes two free bosons and so it possesses
a
chiral $U(1) \times U(1)$ world--sheet
current algebra. The two chiral currents in
question are simply $\partial (\rho e^{\pm i \varphi})$ in polar
coordinates.
In the dual formulation they become non--locally realized and assume
the form
\be
{\Psi}_{\pm} = \partial \left(\rho e^{\pm i \int {1 \over {\rho}^{2}}
(d z \partial \varphi - d \bar{z} \bar{\partial} \varphi)} \right) ~,
\ee
using eq. (A.5). Since they are chirally conserved,
$\bar{\partial} {\Psi}_{\pm} = 0$, they are the parafermions of the
dual model substituting the original $U(1) \times U(1)$
local currents. The parafermions
of opposite chirality may be introduced in a similar way. The vector
fields $J_{\pm}$ are also non--local in the dual formulation, but
they do not deserve the name parafermions, because they are not
chirally
conserved.

\newpage

\centerline{\bf REFERENCES}
\begin{enumerate}
\item A. Giveon, M. Porrati and E. Rabinovici, Phys. Rep.
\underline{244} (1994) 77.
\item A. Font, L. Ibanez, D. Lust and F. Quevedo, Phys. Lett.
\underline{B249} (1990) 35.
\item J. Schwarz and A. Sen, Phys. Lett. \underline{B312} (1993) 105;
A. Sen, Int. J. Mod. Phys. \underline{A9} (1994) 3707.
\item I. Bakas, Nucl. Phys. \underline{B428} (1994) 374;
{\em ``String Dualities and the Geroch Group"}, in the proceedings of
the
satellite colloquium {\em ``Topology, Strings and Integrable Models"}
to the
XIth International Congress of Mathematical Physics, Paris, CERN
preprint
CERN--TH.7499/94, hep--th/9411118, November 1994.
\item A. Sen, Nucl. Phys. \underline{B434} (1995) 179.
\item I. Bakas, Phys. Lett. \underline{B343} (1995) 103.
\item E. Bergshoeff, R. Kallosh and T. Ortin, {\em ``Duality Versus
Supersymmetry and Compactification"}, Groningen preprint UG--8--94,
hep--th/9410230, October 1994.
\item E. Kiritsis, Nucl. Phys. \underline{B405} (1993) 109;
S. Hassan and A. Sen, Nucl. Phys. \underline{B405} (1993) 143;
M. Henningson and C. Nappi, Phys. Rev. \underline{D48} (1993) 861;
A. Giveon and E. Kiritsis, Nucl. Phys. \underline{B411} (1994) 487.
\item S. Hassan, {\em ``O(D,D;R) Deformations of Complex Structures
and
Extended World Sheet Supersymmetry"}, Tata preprint TIFR--TH--94--26,
hep--th/9408060, August 1994.
\item M. Rocek and E. Verlinde, Nucl. Phys. \underline{B373} (1992)
630;
E. Kiritsis, C. Kounnas and D. Lust,
Int. J. Mod. Phys. \underline{A9} (1994) 1361;
I. Ivanov, B. Kim and M. Rocek, Phys. Lett. \underline{B343} (1995)
133.
\item L. Alvarez--Gaume and D. Freedman, Commun. Math. Phys.
\underline{80} (1981) 443;
S. Gates, C. Hull and M. Rocek, Nucl. Phys. \underline{B248} (1984)
157;
P. Howe and G. Papadopoulos, Nucl. Phys. \underline{B289} (1987) 264;
Class. Quant. Grav. \underline{5} (1988) 1647; P. van Nieuwenhuizen
and B. de Wit, Nucl. Phys. \underline{B312} (1989) 58.
\item C. Callan, J. Harvey and A. Strominger, Nucl. Phys.
\underline{B359} (1991) 611.
\item C. Kounnas, Phys. Lett. \underline{B321} (1994) 26;
I. Antoniadis, S. Ferrara and C. Kounnas, Nucl. Phys.
\underline{B421} (1994) 343.
\item A. Giveon, E. Rabinovici and G. Veneziano, Nucl. Phys.
\underline{B322} (1989) 167;
K. Meissner and G. Veneziano, Phys. Lett. \underline{B267} (1991) 33.
\item E. Alvarez, L. Alvarez--Gaume and Y. Lozano, Phys. Lett.
\underline{B336} (1994) 183; {\em ``An Introduction to T--Duality
in String Theory"}, CERN preprint CERN--TH.7486/94, hep--th/9410237,
October 1994.
\item C. Kounnas, M. Porrati and B. Rostant, Phys. Lett.
\underline{B258} (1991) 61.
\item C. Boyer and J. Finley, J. Math. Phys. \underline{23} (1982)
1126;
J. Gegenberg and A. Das, Gen. Rel. Grav. \underline{16} (1984) 817.
\item S. Hawking, Phys. Lett. \underline{A60} (1977) 81;
G. Gibbons and S. Hawking, Commun. Math. Phys. \underline{66}
(1979) 291;
K. Tod and R. Ward, Proc. R. Soc. Lond. \underline{A368} (1979) 411.
\item T. Eguchi, P. Gilkey and A. Hanson, Phys. Rep.
\underline{66} (1980) 213.
\item G. Gibbons and P. Ruback, Commun. Math. Phys. \underline{115}
(1988) 267.
\item I. Bakas, in the proceedings of the Trieste conference
{\em ``Supermembranes and Physics in 2+1 Dimensions"}, eds. M. Duff,
C. Pope and E. Sezgin, World Scientific, Singapore, 1990;
Q.--H. Park, Phys. Lett. \underline{B236} (1990) 429.
\item M. Atiyah and N. Hitchin, Phys. Lett. \underline{A107} (1985)
21;
{\em ``The Geometry and Dynamics of Magnetic Monopoles"},
Princeton University Press, 1988.
\item I. Bakas, work in progress.
\item C. Hull, Phys. Lett. \underline{B178} (1986) 357.
\item A. Dabholkar, G. Gibbons, J. Harvey and F. Ruiz--Ruiz, Nucl.
Phys.
\underline{B340} (1990) 33; A. Strominger, Nucl. Phys.
\underline{B343}
(1990) 167; S. J. Rey, Phys. Rev. \underline{D43} (1991) 526; M. Duff
and X. Lu, Nucl. Phys. \underline{B354} (1991) 141; R. Khuri,
Phys. Lett. \underline{B259} (1991) 261; Nucl. Phys.
\underline{B387} (1992) 315;
M. Bianchi, F. Fucito,
G. Rossi and M. Martellini, {\em ``ALE Instantons in String Effective
Theory"}, Rome preprint ROM2F--94--17, hep--th/9409037, September
1994.
\item A. Sevrin, W. Troost and A. van Proeyen, Phys. Lett.
\underline{B208} (1988) 447.
\item M. Rocek, K. Schoutens and A. Sevrin, Phys. Lett.
\underline{B265} (1991) 303.
\item L. Alvarez--Gaume, E. Alvarez and Y. Lozano, Nucl. Phys.
\underline{B424} (1994) 155.
\end{enumerate}
\end{document}